\documentclass[prl,twocolumn,superscriptaddress,floatfix,showpacs]{revtex4-1}
\usepackage{graphicx}
\usepackage{dcolumn}
\usepackage{bm}
\usepackage{epsf}
\usepackage{subfigure}
\usepackage{epstopdf}
\usepackage{amsmath}
\usepackage{amssymb}

\begin{document}

\title{Stochastically driven single level quantum dot: a nano-scale finite-time thermodynamic machine and its various operational modes}

\author{Massimiliano Esposito}
\affiliation{Complex Systems and Statistical Mechanics, University of Luxembourg, L-1511 Luxembourg,
Luxembourg}
\author{Niraj Kumar}
\author{Katja Lindenberg}
\affiliation{Department of Chemistry and Biochemistry and BioCircuits Institute, University of California San Diego, 9500 Gilman Drive, La Jolla, CA 92093-0340, USA}
\author{Christian Van den Broeck}
\affiliation{Hasselt University, B-3590 Diepenbeek, Belgium}

\pacs{05.70.Ln,05.40.-a,05.20.-y}

\begin{abstract}
We describe a single-level quantum dot in contact with two leads as a nanoscale finite-time thermodynamic machine. The dot is driven by an external stochastic force that switches its energy between two values. In the isothermal regime, it can operate as a rechargeable battery by generating an electric current against the applied bias in response to the stochastic driving, and re-delivering work in the reverse cycle. This behavior is reminiscent of the Parrondo paradox. If there is a thermal gradient the device can function as a work-generating thermal engine, or as a refrigerator that extracts heat from the cold reservoir via the work input of the stochastic driving. The efficiency of the machine at maximum power output is investigated for each mode of operation, and universal features are identified.
\end{abstract}   

\maketitle

Parallel to spectacular developments in bio- and nano-technology, there has been great theoretical interest in the study of small-scale machines. A well documented case is the small-scale Carnot engine, in which the operational unit is subject to thermal fluctuations \cite{ssc, espositoEPL2009, esposito2009, esposito2010}. Of greater biological relevance are machines that convert one form of work to another, and yet these have received far less attention \cite{Isotherm}. In this letter we introduce an electronic nano-device that allows several modes of operation. The device is a single-level quantum dot subject to stochastic driving while in contact with two reservoirs that may be at different temperatures and chemical potentials. Its properties can be derived from a stochastic thermodynamic description \cite{st}. We investigate in analytic detail various operational regimes. When operating under tight coupling conditions, familiar features are recovered in appropriate limits: Carnot efficiency for reversible operation when the reservoirs are at different temperatures, universal features of efficiency at maximum power \cite{VandenBroeck05,esposito2009}, and efficiency at maximum power close to the Curzon-Ahlborn efficiency \cite{curzon}. When the reservoirs are at the same temperature, the work done on the dot by the switching can reverse the ``normal" direction (from high to low chemical potential) of the current. Thus, the engine can be seen as a technologically relevant implementation of the Parrondo paradox \cite{parrondo1} in that the switching can induce an electron flow {\it against} the chemical gradient. When operating under tight coupling, the efficiency at maximum power starts from the universal value of $1/2$ close to equilibrium and increases monotonically to $1$ as one moves further into the nonequilibrium regime, and can thus be much higher than in the traditional implementations of the Parrondo paradox \cite{parrondo2}. The same efficiency is observed when the engine works in the reverse mode.
\begin{figure}[ht]
\vspace{0.5cm}
\rotatebox{0}{\scalebox{0.55}{\includegraphics{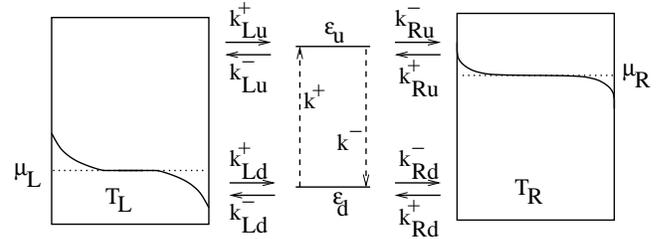}}}
\caption{The model consists of a single level quantum dot. A stochastic external force causes
the energy of the dot to undergo transitions between two values,  
$\varepsilon_u \geq \varepsilon_d$, at random times with rates $k^+$ and $k^{-}$.
The dot also exchanges electrons with two leads that may be at different chemical potentials
$\mu_L$ and $\mu_L$ and temperatures $T_L$ and $T_R$.}
\label{fig:model}
\vspace{-0.3cm}
\end{figure}


{\bf Model and dynamics} - We consider a single-level quantum dot whose energy is stochastically switched between an upper and a lower value, $\varepsilon_j$ with $j=u,d$. The upward and downward rates are $k^{+}$ and $k^{-}$. The dot is in contact with a left and a right lead, $\nu=L,R$, at chemical potentials $\mu_\nu$ and temperatures $T_{\nu}$. The transition rates of an electron into lead $\nu$ from the dot, and out of lead $\nu$ to the dot, are given respectively by $k^{-}_{\nu j}=\Gamma_{\nu j}(1-f_{\nu j})$ and $k^+_{\nu j} =\Gamma_{\nu j}f_{\nu j}$. Here $f_{\nu j}=(1+\exp{\{(\varepsilon_j-\mu_{\nu})/T_{\nu}}\})^{-1}$ is the Fermi distribution in lead $\nu$, and $\Gamma_{\nu j}$ is the coupling strength between this lead  and the dot in state $j$. The four possible states of the system are denoted by $\{u1,u0,d1,d0\}$, where $jn$ defines whether the level $j=u,d$ is empty or occupied, $n=0,1$. The Markovian master equation for the evolution of the  state occupation probabilities in terms of $k^\pm_j \equiv k^\pm_{Lj} +k^\pm_{Rj}$, the total transition rate out of the dot (+) or into the dot (-) from either lead, is given by
\begin{widetext}
\begin{equation}{\label{e5}}
\begin{bmatrix}
\dot{P}_{u1}\\
\dot{P}_{u0}\\
\dot{P}_{d1}\\
\dot{P}_{d0}
\end{bmatrix}
=
\begin{bmatrix}
-(k^{-}+k^-_u) & k^+_u& k^{+} & 0 \\
k^{-}_u& -(k^{-}+k^{+}_u) & 0 & k^{+} \\
k^{-} & 0 & -(k^{+}+k^{-}_d)& k^{+}_d \\
0 & k^{-} & k^{-}_d& -(k^{+}+k^{+}_d)
\end{bmatrix}
\begin{bmatrix}
P_{u1}\\
P_{u0}\\
P_{d1}\\
P_{d0}
\end{bmatrix}.
\end{equation}
\end{widetext}
The probability currents between the four states are $\mathcal I_{un\leftarrow dn} = k^{+}P_{dn}-k^{-}P_{un}$ for $n=0,1$, and $\mathcal I_{j1\leftarrow j0}^{(\nu)} = k^{+}_{\nu j} P_{j0}-k^{-}_{\nu j} P_{j1}$ for $j=u,d$. It is convenient to introduce the total current to the dot when it is down or up, that is, the sum $\mathcal I_{j1\leftarrow j0} = \sum_{\nu} \mathcal I_{j1\leftarrow j0}^{(\nu)}$. Using (\ref{e5}) we easily verify that at steady state there is the appropriate balance between the currents, that is, $\mathcal I \equiv \mathcal I_{u1\leftarrow d1} = -\mathcal I_{u0\leftarrow d0} = -\mathcal I_{u1\leftarrow u0} = \mathcal I_{d1\leftarrow d0}$. The steady state probabilities can be obtained analytically (not shown here) and lead to the following expressions for the currents:
\begin{eqnarray}
&&\mathcal{I} = \left( \frac{k^{+} k^{-}}{k^{+}+k^{-}} \right)
\frac{ \sum_{\nu} \Gamma_{\nu u} \big(\Gamma_{Ld} f_{Ld}+\Gamma_{Rd} f_{Rd}-\Gamma_{d}f_{\nu u} \big)}
{ k^{-} \Gamma_{d} + k^{+} \Gamma_{u} + \Gamma_{d}\Gamma_{u}}, \nonumber \\
&&\mathcal I_{d1\leftarrow d0}^{(L)}  = \left( \frac{k^{-}\Gamma_{Ld}}{k^{+}+k^{-}} \right) \left[  
\frac{(k^{-}+\Gamma_u) \Gamma_{Rd} (f_{Ld}-f_{Rd}) }{ k^{-} \Gamma_{d} + k^{+} (\Gamma_u + \Gamma_d\Gamma_u)} \right. \nonumber \\
&&\hspace{2.4cm} \left. +\frac{k^{+} \left( \Gamma_uf_{Ld}-\Gamma_{Lu} f_{Lu}-\Gamma_{Ru}f_{Ru} \right)}{k^{-} \Gamma_{d} + k^{+} (\Gamma_u + \Gamma_d\Gamma_u)} \right] .\label{AnaCurr}
\end{eqnarray}
We have introduced the combination $\Gamma_d \equiv \Gamma_{Ld} + \Gamma_{Rd}$, and similarly for $\Gamma_u$. $\mathcal I_{u1\leftarrow u0}^{(L)}$ is obtained from $\mathcal I_{d1\leftarrow d0}^{(L)}$ with the substitutions $u \leftrightarrow d$ and $k^+ \leftrightarrow k^-$.


{\bf Thermodynamics} - The energy current injected in the system by the stochastic driving reads
\begin{equation}{\label{e11}}
\mathcal I_{ext} = (\varepsilon_u - \varepsilon_d) \mathcal I_{u1\leftarrow d1} = (\varepsilon_u - \varepsilon_d) \mathcal I,
\end{equation}
while the matter ($M$) and energy ($E$) currents entering the system from lead $\nu$ are given by 
\begin{eqnarray}{\label{E&Mcurr}}
\mathcal I_M^{(\nu)}=\displaystyle\sum_{j} \mathcal I_{j1\leftarrow j0}^{(\nu)} \ \ , \ \
\mathcal I_E^{(\nu)}=\displaystyle\sum_{j} \varepsilon_j \mathcal I_{j1\leftarrow j0}^{(\nu)} .
\end{eqnarray}
The heat flux from the lead $\nu$ is  
\begin{equation}{\label{e19}}
\dot Q^{(\nu)}=  \mathcal I_E^{(\nu)} - \mu_{\nu} \mathcal I_M^{(\nu)}.
\end{equation}
It is easy to verify matter and energy conservation in the steady state, $\mathcal I_M^{(L)}=-\mathcal I_M^{(R)}$ and $\mathcal I_{ext} = - \mathcal I_E^{(L)} - \mathcal I_E^{(R)}$. As a result, power becomes the sum of two contributions,
\begin{equation}{\label{WorkCurr}}
\dot{W} = - \sum_{\nu} \dot Q^{(\nu)} = (\varepsilon_u - \varepsilon_d) \mathcal I +
(\mu_{R}-\mu_{L}) \mathcal I_M^{(L)}.
\end{equation}
The first is the contribution of the energy flux injected by the stochastic driving. The second is the energy flux required to bring an electron from the left lead through the dot to the right lead. Since at steady state entropy production, $\dot S_i$, is minus the entropy flow, that is, $\dot S_i=-\sum_{\nu} \dot Q^{(\nu)}/T_{\nu}\geq 0$, we find that entropy production is the sum
of three force-flux terms,
\begin{eqnarray}{\label{EP}}
\hspace{-0.3cm} \dot S_i = \frac{(\varepsilon_u - \varepsilon_d)}{T_R} \mathcal{I} + (\frac{\mu_L}{T_L}-\frac{\mu_R}{T_R}) \mathcal{I}_M^{(L)} + (\frac{1}{T_R}-\frac{1}{T_L}) \mathcal{I}_E^{(L)} . \;
\end{eqnarray}

The system reaches equilibrium when entropy production vanish, $\dot S_i = 0$, which implies that all currents in the system also vanish. In general, this requires that the three thermodynamic forces vanish separately, i.e., that $\varepsilon_u = \varepsilon_d$, $\mu_L=\mu_R$ and $T_L=T_R$. This is however not necessary when $\Gamma_{Lu}/\Gamma_{Ld} \to 0$ (disallowing transitions from $u$ to $L$), and $\Gamma_{Rd}/\Gamma_{Ru} \to 0$ (disallowing transitions from $d$ to $R$). This combined limit corresponds to a regime of {\it tight coupling}, where the transport of matter, energy and heat become proportional to each other because, by removing the possibility of transitions from $u$ to $L$ and from $d$ to $R$, there is a single pathway connecting the left and right leads. As a result entropy production can be expressed as $\dot S_i = X \mathcal{I}$, i.e. in terms of the ``collapsed" effective force
\begin{eqnarray}{\label{ForceSC}}
X= \frac{(\varepsilon_u - \varepsilon_d)}{T_R} + (\frac{\mu_L}{T_L}-\frac{\mu_R}{T_R}) + \varepsilon_d (\frac{1}{T_R}-\frac{1}{T_L}),
\end{eqnarray}
and the single flux $\mathcal{I}$
\begin{eqnarray}{\label{CurrStrCoup}}
\mathcal{I} = \mathcal{I}_{M}^{(L)} = \mathcal{I}_{E}^{(L)}/\varepsilon_d = \alpha (f_{Ld} - f_{Ru}) ,
\end{eqnarray}
where $\alpha = (k^{+} k^{-} \Gamma_{Ld} \Gamma_{Ru})/\{(k^{+}+k^{-})(k^{-}\Gamma_{Ld}+k^{+}\Gamma_{Ru}+\Gamma_{Ld}\Gamma_{Ru})\}$. We now see that equilibrium only requires the effective force to vanish, $X=0$, without the requirement that the three thermodynamic forces vanish separately.


{\bf Current rectifier and Parrondo paradox} - We return to the general expression (\ref{EP}) and take the two leads to be at the same temperature, $T=T_L=T_R$ but at different chemical potentials. We define $\Delta \mu = \mu_{R} - \mu_{L} \geq 0$ and $\Delta \varepsilon = \varepsilon_u - \varepsilon_d  \geq 0$. Entropy production then becomes 
\begin{eqnarray}{\label{EPrecti}}
T \dot S_i = \Delta \varepsilon \; \mathcal{I} - \Delta \mu \; \mathcal{I}_M^{(L)} \geq 0.
\end{eqnarray}
In the absence of the external (stochastic) driving, the quantum dot remains in its initial state, either in state $u$ or in state $d$. As a result of the direction of the chemical gradient, the electronic current will be negative whichever state the dot is in, that is, $\mathcal I_{j1 \leftarrow j0}^{(L)} <0$ for $j=u,d$. However, when the stochastic driving induces switching between the two states, the resulting net current can invert and become positive, with electrons flowing {\it against the chemical bias}. This remarkable effect can be seen as a new version of the Parrondo paradox \cite{parrondo1}. Thermodynamically, the phenomenon corresponds to the transformation of one type of work, that which is involved in the modulation of the dot energy level, into another, the pumping of electrons from low to high chemical potential. The power at which this process takes place, and the corresponding efficiency, are given by: 
\begin{eqnarray}{\label{power}}
{\cal P} = \Delta \mu \; \mathcal{I}_M^L , \quad  0 \leq \eta = \frac{\Delta \mu \; \mathcal{I}_M^L}{\Delta \varepsilon \; \mathcal{I}} \leq 1.
\end{eqnarray}

We carry the analysis further in the tight coupling limit, $\mathcal{I}_M^L = \mathcal{I}$, where the efficiency $\eta$ becomes current independent, $\eta = \Delta \mu/\Delta \varepsilon$. The two forces appearing in Eq.~(\ref{EP}), $\Delta \varepsilon/T$ and $\Delta \mu/T$, collapse into a single one, $X =(\Delta \varepsilon-\Delta \mu)/T$. Furthermore, from (\ref{ForceSC}) we see that the condition for pumping electrons against the bias, $f_{Ld} > f_{Ru}$, is achieved with the Fermi distributions when $\Delta\varepsilon -\Delta\mu >0$, that is, when $X>0$. The upper bound of the efficiency (\ref{power}) is $\eta=1$ (when $\Delta \mu = \Delta \varepsilon$) and is reached when entropy production vanishes, $X=0$. This, however implies that the current and hence the power vanishes. This reversible operation is not the useful limit to consider in practice. Instead, we next consider the efficiency at maximum power. 

We start by analyzing the linear response regime near equilibrium, where currents are expanded to first order in the forces, $\mathcal{I} = L_{11} \Delta \varepsilon - L_{12} \Delta \mu$ and $\mathcal{I}_M^L= L_{21} \Delta \varepsilon - L_{22} \Delta \mu$. The Onsager coefficients can be calculated
analytically, and the reciprocity relation $L_{12}=L_{21}$ can then easily be verified. Maximizing the power output ${\cal P}$ with respect to the chemical bias $\Delta \mu$ leads to the condition $\Delta \mu^*=\Delta \varepsilon L_{12}/(2L_{22})$, which in turn yields the well-known result for the efficiency at maximum power in the linear regime \cite{VandenBroeck05},
\begin{eqnarray}{\label{EffMaxP}}
\eta^* = \frac{L_{12}^2}{2L_{11}L_{22}+2(L_{11}L_{22}-L_{12}^2)} \leq \frac{1}{2}.
\end{eqnarray}
Tight coupling in the linear regime implies the usual relations among the Onsager coefficients, $L_{11}=L_{22}=L_{12}$. In this case the efficiency at maximum power (\ref{EffMaxP}) reaches its upper bound, $\eta^*=1/2$.

\begin{figure}
\rotatebox{0}{\scalebox{1.2}{\includegraphics{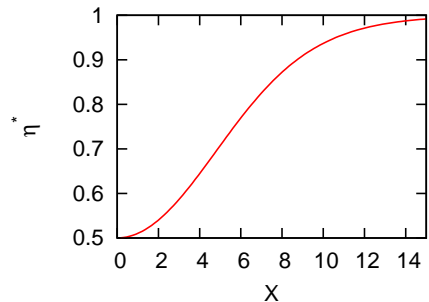}}}
\vspace{-0.5cm}
\caption{Efficiency at maximum power as a function of the thermodynamic force $X$ [cf. \ref{FinalEffRectif}].}
\label{FigConverter}
\end{figure}

To go beyond the linear regime, we assume tight coupling from the outset and introduce the convenient combinations $x_L=(\varepsilon_d-\mu_L)/T$ and $x_R=(\varepsilon_u-\mu_R)/T$, in terms of which the Fermi distribution is $f(x)=(1+\exp{x})^{-1}$. Using (\ref{ForceSC}) in (\ref{power}), power can be rewritten as ${\cal P}  = \alpha \left(T(x_L-x_R)+\Delta \varepsilon \right) \left( f(x_L)-f(x_R) \right)$. Maximizing power with respect to $x_L$ and $x_R$ leads to $x_R=-x_L$ and $x_R+\sinh x_R=\Delta \varepsilon/2T$. Using these results and the fact that $X=x_R-x_L$, the efficiency at maximum power is found to be  
\begin{equation}{\label{FinalEffRectif}}
\eta^\star=1-\frac{X}{X+2 \sinh \left(\frac{X}{2 }\right)}.
\end{equation}
Expanding this result close to equilibrium, we find: $\eta^\star=1/2+X^2/96-X^4/11520+O\left(X^6\right)$. Remarkably, the efficiency at maximum power increases monotonically from the linear regime value $\eta^\star=1/2$ ($X \to 0$) to $\eta^\star=1$, reached when $X \to \infty$, cf. Fig. \ref{FigConverter}.

The above calculations can be repeated when the engine operates in reverse, using the difference in chemical potential as input work and the modulation of the energy level as output. This leads to 
\begin{eqnarray}{\label{powerbis}}
{\cal P} = \Delta \varepsilon \; \mathcal{I} , \quad 
0 \leq \eta = \frac{\Delta \varepsilon \; \mathcal{I}}{\Delta \mu \; \mathcal{I}_M^L} \leq 1,
\end{eqnarray}
Proceeding along the same lines as before, we find that the resulting efficiency at maximum power in the tight coupling regime is again given by Eq. (\ref{FinalEffRectif}). Hence both modes of operation, forward and backward, can have very high efficiency, suggesting a possible technological interest of the device.


{\bf Refrigerator and heat engine} - While the isothermal operation of our engine as described above is its most interesting feature, it is instructive to ascertain that it reproduces known behavior under more conventional operating conditions, namely, when the leads are at different temperatures, say $T_{L} \geq T_{R}$. For this recovery it is sufficient to consider the simplified case of equal chemical potentials, $\mu=\mu_L=\mu_R$. Entropy production now reads
\begin{eqnarray}{\label{EPref}}
T_R \dot S_i = \Delta \varepsilon \mathcal{I} + \eta_C \dot{Q}^{(L)} \geq 0,
\end{eqnarray}
where $\eta_C=1-T_R/T_L$ is the Carnot efficiency. Our device operates as a refrigerator when the external driving extracts heat from the cold reservoir, $\dot{Q}^{(R)} \geq 0$. The power output and coefficient of performance $\bar{\eta}$ of this process are given by
\begin{eqnarray}{\label{COP}}
{\cal P} = \dot{Q}^{(R)} , \quad  0 \leq \bar{\eta} = \frac{\dot{Q}^{(R)}}{\Delta \varepsilon \;
\mathcal{I}} \leq \frac{1}{\eta_C}-1.
\end{eqnarray}
When functioning as a heat engine, the machine produces net work on the stochastic driving process, i.e. $\mathcal{I}_{ext} \leq 0$, at the cost of a driving heat flow $\dot{Q}^{(L)} \leq 0$. The power output and efficiency of this transformation are given by    
\begin{eqnarray}{\label{EffThermalEng}}
{\cal P} = -\Delta \varepsilon \; \mathcal{I} , \quad 0 \leq \eta = \frac{- \Delta \varepsilon \; \mathcal{I}}{\dot{Q}^{(L)}} \leq \eta_C 
\end{eqnarray}

In the tight coupling limit, the collapsed force (\ref{ForceSC}) appearing in the entropy production, $\dot S_i = X \mathcal{I}$, becomes $X=x_R-x_L$ with $x_R=(\varepsilon_u-\mu)/T_R$ and $x_L=(\varepsilon_d-\mu)/T_L$. Here we have used $\mathcal I_M^{(L)}=\mathcal I$ and $\mathcal I_E^{(L)}=\varepsilon_d \mathcal I$. The efficiencies (\ref{COP}) and (\ref{EffThermalEng}) now reduce to
\begin{eqnarray}{\label{EffEngineSC}}
&&\bar{\eta}=\frac{\mu-\varepsilon_u}{\Delta \varepsilon}=\frac{x_R(1-\eta_C)}{x_L-(1-\eta_C)x_R},\nonumber\\
&&\eta=-\frac{\Delta \varepsilon}{\varepsilon_d-\mu}=1-\frac{x_R}{x_L}(1-\eta_C).
\end{eqnarray}

\begin{figure}
\rotatebox{0}{\scalebox{1.5}{\includegraphics{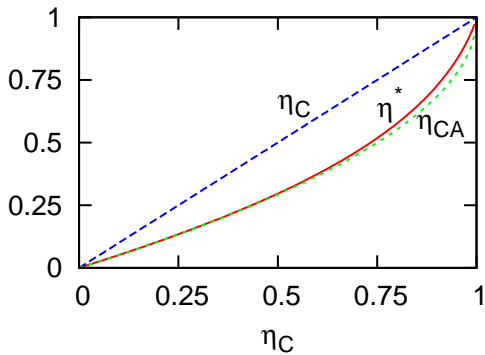}}}
\vspace{-0.8cm}
\caption{Efficiency at maximum power, $\eta^\star$, for the thermal engine as a function of Carnot efficiency $\eta_C$, in comparison with the Curzon-Ahlborn efficiency $\eta_{CA}=1-\sqrt{1-\eta_C}$.}
\label{fig:heat}
\end{figure}

Turning to the regime of maximum power in the tight coupling regime, we first discuss the heat engine. In order to maximize the output power, $\mathcal P= -(\varepsilon_u-\varepsilon_d)\mathcal I=T_L\left(x_L-x_R(1-\eta_C)\right)\mathcal I(x_L,x_R)$, with respect to $x_L$ and $x_R$, we need to solve $\partial \mathcal P/\partial x_L=0,~\partial \mathcal P/\partial x_R=0$. The procedure is identical to that of Ref. \cite{espositoEPL2009}. The resulting transcendental equation can easily be solved numerically and leads to the efficiency at maximum power displayed in Fig.~\ref{fig:heat}. We note that $\eta^\star$ increases monotonically when driven out of equilibrium. It is bounded from above by $\eta_C$, while the Curzon-Ahlborn efficiency $\eta_{CA}=1-\sqrt{1-\eta_C}$ provides a rather tight lower bound. The transcendental equation can also be solved perturbatively for small $\eta_C$, 
\begin{equation}
\eta^\star=\frac{\eta_C}{2}+\frac{\eta_C^2}{8}+O(\eta_C^3).
\end{equation}
We thus recover the universal value $\eta_C/2$ in the linear regime \cite{VandenBroeck05}, as well as the factor $1/8$ for the quadratic coefficient. This latter result thus again  supports the universality of this value (for systems with a left/right symmetry) \cite{esposito2009}.

Finally, we turn to the efficiency at maximum power for the refrigerator in the tight coupling regime. The output power $\dot Q^{(R)}=-(X+x_L)T_R \mathcal{I}(X,x_L)$ has a local maximum with respect to the thermodynamic force $X$ which cannot be found analytically. We can solve $\partial \dot Q^{(R)}/\partial X=0$ perturbatively as a power series in $X$ by expanding $x_L=a_0+a_1X+a_2X^2+a_3X^3+a_4 X^4 + O(X^5)$. We find $a_0=a_2=a_4=0$, $a_1=-2$, $a_3=1/3$. Inserting the result in Eq. (\ref{COP}) leads to
\begin{equation}
\bar{\eta}^\star=\frac{\bar{\eta}_{\rm id}}{\bar{\eta}_{\rm id}+2}-\frac{\left(\bar{\eta}_{\rm id}^2
+\bar{\eta}_{\rm id}\right) X^2}{3 (\bar{\eta}_{\rm id}+2)^2}+O\left(X^4\right),
\end{equation}
where $\bar{\eta}_{\rm id}=\eta_C^{-1}-1$ is the efficiency in the reversible limit.

{\bf Discussion} - We have presented a detailed analysis of a stochastically driven single-level electronic nano-device. When operating as a thermal engine or a refrigerator, our model reproduces all the expected results. However, of special interest is the isothermal case, where the device can be used as a work to work converter. It can be seen as a novel implementation of the Parrondo paradox, with electrons moving up in chemical potential under the influence of the randomly switching energy level. The asymmetry in the system is realized via the tight coupling condition, which implies that each of the energy levels of the dot is coupled to a single heat bath. In this case, the efficiency at maximum power is very high (up to $1$), suggesting the potential technological importance of this mode of operation. Remarkably, the efficiency of the device is equally high in the reverse mode, where work is extracted from electrons moving down in chemical potential. Hence, our device can alternate, for example, between conversion of chemical energy into electrical energy, and vice-versa, thus operating as a highly efficient rechargeable battery. It remains to be seen whether the simplifications that allow the detailed analysis presented here, such as the fully asymmetric coupling, the absence of line broadening, and the weak coupling assumption do not significantly reduce the device efficiency, and whether the technological challenges that the operation of such a nano-scale device present, can be overcome. 


M. E. is supported by the National Research Fund, Luxembourg in the frame of project FNR/A11/02.
This research is supported in part by the NSF under Grant No. PHY-0855471.

\vspace{-0.4cm}



\end{document}